\documentstyle[12pt]{article}
\newcommand{\beq}{\begin{equation}}
\newcommand{\eeq}{\end{equation}}

\newcommand{\dd}{D\hspace{-.65em}/}
\newcommand{\nn}{\partial\hspace{-.55em}/}

\newcommand{\r}{\varphi}

\newcommand{\e}{\bar{\eta}}
\newcommand{\p}{\bar{\psi}}
\newcommand{\s}{\varepsilon}

\def\op{operator}
\def\dop{Dirac operator}
\def\tly{topologically}
\def\tl{topological}
\def\exl{external}
\def\inst{instanton}
\def\sp{sphaleron}
\def\gl{global}
\def\an{anomaly}
\def\f{field}
\def\exst{existence}
\def\nl{normalizable}
\def\nty{normalizability}
\def\zm{zero mode}
\def\lc{level crossing}
\def\mm{mass matrix}
\def\igty{integrability}
\def\cn{condition}
\def\con{configuration}
\begin{document}
\begin{titlepage}
\begin{flushright}
NBI-HE-93-46  \\
August 1993\\
\end{flushright}
\vspace{0.5cm}
\begin{center}
{\large {\bf The Level Crossing Phenomenon with Yukawa Interactions}}\\
\vspace{1.5cm}
{\bf Minos Axenides}
\footnote{e-mail:axenides@nbivax.nbi.dk}\\
\vspace{0.4cm}
{\em The Niels Bohr Institute\\
University of Copenhagen, 17 Blegdamsvej, 2100 Copenhagen, Denmark}\\
\vspace{0.4cm}
{\bf Andrei Johansen}
\footnote{e-mail:johansen@lnpi.spb.su}\\
\vspace{0.4cm}
{\em The St.Petersburg Nuclear Physics Institute\\
 Gatchina, St.Petersburg District, 188350 Russia}\\
\vspace{0.4cm}
{\bf Holger Bech Nielsen}
\footnote{e-mail:hbech@nbivax.nbi.dk}\\
\vspace{0.4cm}
{\em The Niels Bohr Institute\\
University of Copenhagen, 17 Blegdamsvej, 2100 Copenhagen, Denmark}\\

\end{center}
\begin{abstract}
In the minimal model of electroweak interactions we carefully
investigate the spectrum of the massive euclidean \dop \ in three,
four and five dimensions ($D$) in the presence of \tly \ nontrivial
\exl \ fields. More specifically we study the cases of the \inst \
($D=4$),
\sp \ ($D=3$) as well as that of
the ($D=5$) \dop \ that pertains to the existence
of the \gl \ $SU(2)$ \an .
We establish the \exst \ of \nl \ massive fermion \zm s in
all three cases. We
give closed form expressions which relate the massive with the massless
\zm s. As a consequence the \lc \ phenomenon is shown to be manifest and
generic in the presence of Yukawa interactions.
\end{abstract}
\end{titlepage}
\newpage

\section{Introduction}
\setcounter{equation}{0}

The phenomenon of fermion level crossing and its relation to
the existence of normalizable fermionic zero modes of the \dop \ in
the presence of \tly \ non-trivial gauge fields is well known \cite{sc}.
It has been extensively studied and firmly established for the case of
massless fermions.
To be more precise let us discuss it in three superficially different
physical circumstances. Firstly, the \nl \
\zm \ of the $D=4$ euclidean \dop \ in the weak \inst \ \f .
Secondly, the \nl \ solution of the zero energy Dirac equation in the
\sp \ \f \ and thirdly the \gl \ \an \ for an $SU(2)$ gauge theory
coupled to an odd number of Weyl fermion doublets.
The first two examples are related to the problem of baryon number
violation in the standard electro-weak model \cite{hooft}.
In this case the baryon ($B$) and lepton ($L$) number violations are
given by
\beq
\Delta B = \Delta L = - n_g \Delta N_{CS}
\eeq
This effect is due to the fluctuations of the gauge and Higgs fields
over the barrier between topologically inequivalent vacua that correspond to
different Chern-Simons numbers $N_{CS}$. In eq.(1.1) $n_g$
is the number of generations.
More importantly it can be equivalent seen as the fermionic energy level
crossing
in the presence of
the time dependent external
gauge and Higgs fields with a variable Chern-Simons characteristic
\cite{callan,kiskis,christ,amb,rubakov}.
In this process the occupied positive energy level appear
or disappear according to the value of $\Delta N_{CS}$.
In turn the level crossing for fermionic Hamiltonian is equivalent
to the existence of normalizable zero modes of the $D=4$ Dirac \op \
in a euclidean topologically non-trivial external field.
The sphaleron configuration on the other hand is a saddle
point of the static energy
functional which has $N_{CS}=1/2$.
It has been found \cite{boguta,ringwald} that the zero-energy
Dirac equation in the sphaleron field has a normalizable solution
that corresponds exactly to the crossing point of a zero energy level.

The $SU(2)$ global anomaly \cite{witten}
is also related to the phenomenon of
level crossing of the $D=4$ Dirac operator.
In analogy with the case of \inst s\ and \sp s\ this eigenvalue flow
corresponds to the existence of a \zm \ for an appropriately defined
$D=5$ \dop \
in the presence of an external topologically
non-trivial gauge field. As a result the fermionic path integral
is not gauge invariant and the theory is not self-consistent.

The problem that we address in the present work
is common to all the above mentioned physical realizations of the \lc \
phenomenon.
In the realistic electro-weak theory fermions become massive
due to the Higgs mechanism.
In the presence of Yukawa interactions
the Higgs field is a non-vanishing constant at spatial infinity.
In this paper we consider the effect of
non-vanishing Yukawa couplings and fermion masses on the problem
of existence of fermionic zero modes in the presence of
topologically non-trivial
external fields.
While some preliminary work for the \inst \ \cite{rub,anselm} and \sp \
\cite{km,ringwald} cases has already been done we clarify some of the
conceptual issues that arise in these works.
Moreover we attempt to reach a single physical point of view on the
question of existence of massive \zm s\ and \lc \ in the presence of a
fermion mass matrix $M$ for all three cases.

At first it could appear surprising the occurence of \lc \ in the
spectrum of massive fermions. We present a
heuristic physical picture that makes it plausible.
To that end it is sufficient to establish at some point in $\bf R^3,
\bf R^4,
\bf R^5$ respectively vanishing fermion masses with singular external
gauge fields.
For the \inst \ case $(D=4)$ this is trivially true.
In the singular gauge the gauge \f \ falls off to zero at $\bf R^4$ infinity
while it is singular at the \inst \ center where the mass matrix is
zero.
This is also guaranteed by the well-known topological
theorem $\pi^3(SU(2))=Z$. As a result the \zm \ wave function is
concentrated at the center of the \inst .\
Indeed this fact is confirmed by an explicit construction of massive
\zm s\ for some special cases \cite{rub,anselm}.
Though there exists no index theorem for the $D=3$ \sp \ case as
$\pi^2(SU(2))=0$ \cite{hu} there still exists a point in $\bf R^3$ where
fermion masses are zero \cite{km}.

For the case of the $SU(2)$ \an \ ($D=5$) it
is somewhat difficult to demonstrate our claims as there are no explicit
\inst -like configurations for $D=5$.
We may however offer an intuitive argument for the existence of the \lc \
in this case as well.
Let us firstly consider the $D=5$ \dop \
in an external \tly \ nontrivial gauge \f .
Such a configuration is \tly \ guaranteed by the fact that
$\pi^4(SU(2))=Z_2$ \cite{hu}.
In our case where Yukawa interactions are present it is tacitly taken that
Higgs \f s are suitably included as external \f s.
As a consequence of our construction we take at $\bf R^5$ infinity our gauge
\f s to be a pure gauge $A_{\mu}\propto U^{-1}\partial _{\mu}U$ with
$M\propto U$ being the fermion \mm .
Here $U$ is a unitary matrix, a noncontractible map from $S^4$ into
$SU(2)$ on a sphere at $\bf R^5$ infinity.
It means that one can continuously deform it to the identity everywhere
on the sphere except for the region near a single point where $U\neq 1$
(a singularity).
As the radius $R$ of the sphere takes value from infinity to zero we form
a line where the \tl \ non-triviality is concentrated in its
neighbourhood and the external \f \ changes very rapidly.
Therefore at $R=0$ we should have either zero or singularity.
By continuity of the \mm \ the latter is impossible, and hence we should
expect the existence of zero in $M$ with \lc \ to occur.
In the main part of our work we will explicitly demonstrate the
existence of the massive fermion \zm \ along with the spectral flow
of the appropriate \dop s\ for $D=3,4,5$.
In each case we examine their integrability property and give closed
form expressions that relate the \zm \ wave functions for the massive
fermions with those of the massless ones.

The paper is organized as follows. In sect.2 the case of zero modes
for the massive fermions
in the weak instanton field is worked out. Sect.3 is devoted to the case of
a sphaleron. The generalization of Witten's $SU(2)$
anomaly for the case of massive
fermions is carried through in sect.4.
There we prove that the Atiyah-Singer mod 2 index
theorem is sufficient to guarantee the existence of an odd number of zero
modes for the massive $D=5$ \dop \ in the presence of \tly \
non-trivial external \f s.
We finally close with some final comments and speculations on unanswered
questions.

\section{Massive Fermions in a Weak Instanton field}
\setcounter{equation}{0}

The properties of the massive euclidean \dop \ $(D=4)$ in external
weak \inst \ \f s has been considered
in refs.\cite{rub,anselm}. Many of our conclusions are therefore likely
to overlap with those of the above mentioned authors but we include
them here for completeness.
The \inst \ configuration contains both non-trivial configurations of
gauge and Higgs fields.
Let us consider the functional integral over fermionic fields in a
topologically non-trivial  external gauge and Higgs fields.
Though the explicit form of
this configuration
is not essential, in what follows we will make use of
the ones given by t'Hooft \cite{hooft}. For the case
of an anti-instanton field:
\beq
W^a_\mu = \frac{2}{x^2+\rho^2}\;\e_{a\mu\nu}x_\nu.
\eeq
The Higgs field is given by
\beq
\r(x) = -i(\tau^-_\mu x_\mu)\left( \begin{array}{c} 0\\ v/\sqrt{2}
\end{array}  \right)\; \frac{1}{\sqrt{x^2+\rho^2}}.
\eeq
Here  $\rho$ is the scale of the anti-instanton, and $v$ is the
vacuum expectation value of the field $\r^0$.
The t'Hooft's symbols $\e_{abc}$
are defined as $\e_{abc}=\s_{abc}$ with $a,b,c,= 1,2,3$ and
$\e_{a4b} = -\e_{ab4}
=\delta_{ab}, \; \eta_{a44} =0; \; \tau^-_\mu= (\vec{\tau},+i)$. The
expressions (2.1) and (2.2) are given in the regular gauge.

As to the fermions, we consider for simplicity two weak doublets only.
For
instance, these may be
two quark doublets with different colour indices, or one quark and
one lepton doublet. For definiteness they are:
\beq
q_L = \left( \begin{array}{c} u\\ d\end{array}\right)_L, \qquad
q'_L = \left( \begin{array}{c} u'\\ d' \end{array} \right)_L,
\eeq
with associated weak singlet right-handed states as $u_R,\, d_R,\, u'_R,
\,d'_R$.  The generalization to the case of the twelve existing doublets is
trivial.

When the weak doublets are considered in pairs it is possible to pass to the
pure vector interaction of fermions with $W$ bosons. To achieve
this we introduce
instead of fermion fields $q'$ the charge-conjugated fields:
\beq
\tilde{q}'_R =\,\s C\bar{q}'_L = \left( \begin{array}{c} C\bar{d}'_L\\
-C\bar{u}'_L \end{array}\right), \qquad \tilde{q}'_L=\s
Cq'_R=\left(\begin{array}{c} C\bar{d }'_R \\ -C\bar{u}'_R \end{array}\right).
\eeq
Here $\s=i\sigma_2$ acts on the isotopic indices, $C$ is the
charge-conjugation matrix. By introducing
\beq
\begin{array}{lll} \psi=\psi_L+\psi_R,        &  \psi_L=q_L,   &
\psi_R=C\s\bar{q}'_L \\  \eta=\eta_R+\eta_L, & \eta_R=q_R, &
\eta_L=C\s\bar{q}'_R, \end{array}
\eeq
it becomes obvious that both components
of $\psi,\, \psi_L$ and $\psi_R$, are weak doublets while $\eta_R$ and
$\eta_L$ are singlets. Therefore only the $\psi$ field has a vector gauge
interaction:
\beq
L_W=i\p \dd\psi + i\e\nn\eta, \qquad  \dd=\gamma_\mu D_\mu=\gamma_\mu
(\partial_\mu -iW_\mu).
\eeq
Clearly the mixing of the quark and antiquark fields in eq.(2.5) is very
unnatural with respect to colour and electric charge (weak hypercharge).
As those interactions are
irrelevant to our problem at hand from now on we only keep the Yukawa
couplings:
\beq
-L_Y=h_u
\bar{q}_{Li}\s_{ij}u_R\r^*_j+h_d\bar{q}_{Li}d_R\r^i+H.c.+(u,d,h_u,h_d\to
u',d',h'_u,h'_d).
\eeq

Here the Higgs field is $\r^i=(\r^+,\r^0)$ and the fermion masses are
given by $m_u=h_uv/\sqrt{2}, \; m_d=h_dv/\sqrt{2},\; m'_u=h'_uv/\sqrt{2},\;
m'_d=h'_dv/\sqrt{2}$.

By using the fields $\psi$ and $\eta$ of eq.(2.5)
one can rewrite the Lagrangian
(2.7) as follows:
\beq
-L_Y =\p_L M\eta_R +\e_R M^+\psi_L-\p_R\s M'^*\s\eta_L-\e_L\s M'^T
\s\psi_R,
\eeq
where the mass matrix $M(x)$ is given by
\beq
M(x)\; =\; \left( \begin{array}{cc}  h_u\r^0(x)^*,  &   h_d\r^+(x)\\
-h_u\r^+(x)^*,  &  h_d\r^0(x)  \end{array} \right).
\eeq
$M'$ is taken from $M$ by substituting $h_u,h_d\to h'_uh'_d$.

We now make a Euclidean rotation upon which
the fermion fields rotate into
\beq
\psi,\, \eta\; \to\; \psi,\, \eta,   \qquad  \p,\, \e\; \to\; -\,i\psi^+,\,
-\,i\eta^+.
\eeq

The mass terms in $L_Y$ contain now the bilinear
combinations of the fields with the same chirality.
The Lagrangian reads now as follows
\beq
L = L_W + L_Y,
\eeq
\beq
L_W  =  -i\psi^+ \dd\,\psi\, -\, i\eta^+\nn\,\eta ,
\eeq
$$L_Y  =  -i\psi^+_RM\eta_R - i\eta^+_L M^+\psi_L +i\psi^+_L\s M'^*\s\eta_L
+i\eta^+_R\s M'^T\s\psi_R.$$
It is well known that in the absence of the Yukawa couplings a massless fermion
has a zero-mode in the (anti)instanton field.
At first
glance one would expect that when
fermions acquire masses in the presence of the Yukawa couplings
the zero mode disappears. One can
see this  not to be true.
In order to look for the wave function for the fermion zero mode
one should consider the
classical equations of motion for the fields $\psi$ and $\eta$ in the
external gauge and Higgs fields. In the Euclidean version of
the theory they are obtained by a variation of the Lagrangian
and they are given by:

\begin{eqnarray}
\dd\,\psi_L   = \;-M\eta_R, \qquad  \nn\,\eta_R  = \; -M^+\psi_L,\\
\dd\,\psi_R  = \;\s M'^*\s\eta_L, \qquad \nn\,\eta_L  = \;\s
M'^T\s\psi_R.
\end{eqnarray}
It is known
\cite{hooft} that for a massless fermion there is a zero mode of right-handed
chirality in the instanton field as well as
one of a left-handed chirality for
the anti-instanton field (2.1). As it was pointed out in ref.
\cite{anselm}
there exists the generalization of the usual $\gamma_5$-chirality for the
case of massive fermions, namely that of the $\Gamma_5$-chirality.
It is defined to be
$ \Gamma_{5} = (-1)^{2T+1}\gamma_{5}$ where $T$ is the weak
isospin.
The solutions to
eqs.(2.13) and (2.14)
are classified by the\ $\Gamma_5=\pm1$\ eigenvalues.
For massive leptons or quark fields\ $\Gamma_{5}=+1$\ (left handed fermions
have\ $T=1/2$\ while right handed ones\ $T=0$\ ). For antileptons and
antiquarks\ $\Gamma_{5}=-1$\ . The index theorem that relates the number of
left\ ($n_{L}$)\ and right\ ($n_{R}$)\ handed fermion zero modes to the
topological charge of an external gauge field\ $Q_{T}$\ given by\
$n_{L}-n_{R}=-Q_{T}$\ generalizes for the case of massive fermion zero
modes.

In what follows we will present a closed form expression for such
a massive zero mode in the model under consideration. At first let us
consider an anti-instanton configuration. Since for $M=M'=0$ there
is a left-handed $\psi$
zero-mode we expect that\ $\Gamma_5=+1$, i.e. we look for solutions
$\psi_R=0, \;\eta_L=0$ but with $\psi_L\neq0, \; \eta_R\neq0$.
In the instanton field one should solve
eqs.(2.14).
Let us denote the wave function of the massless Dirac operator
as $\psi_{0L}$ with
\beq
\dd\, \psi_{0L} \; =\; 0.
\eeq
We are looking for
a solution to eqs.(2.13) with the following features:

i) it coincides with $\psi_{0L}$ in the $M\to 0$ limit,

ii) it is normalizable.

We try to find such a solution iteratively.
To this end we express $\eta_R$ in terms of $\psi_L$ and substitute it
back to eq.(2.13).
We get
\beq
\dd \psi_L = M \frac{1}{\nn} M^+ \psi_L.
\eeq
If we start with $\psi^{(0)}_L = \psi_{0L}$ as the zero level
approximation one then gets the following equation for the first level
approximation $\psi^{(1)}_L$:
\beq
\dd \psi^{(1)}_L = M \frac{1}{\nn} M^+ \psi _{0L}
\eeq
where $\psi_L = \psi_{0L} + \psi^{(1)}_L+...$ We assume that
$\psi^{(1)}_L$ is orthogonal to $\psi_{0L}$.
At this stage of the iteration we must formally check the integrability
of this equation. This is a necessary condition for the consistency
of the chosen zero level approximation. Indeed the left hand side
of eq.(2.17)
is orthogonal to $\psi_{0L}$ since the \op \ $\dd\;$  is antihermitean
and annihilates the $\psi_{0L}$ outstate
when we integrate it by parts.
Fortunately the integrability \cn \ is automatically satisfied here
because the massless \zm \ wave function $\psi_{0L}$
is a Weyl spinor.
However this is not the case with the \sp \ \f \ and the $SU(2)$ \gl \
\an .
{}From eq.(2.17) we may easily find
\beq
\psi^{(1)}_L = \frac{1}{\dd} M \frac{1}{\nn} M^+ \psi_{0L}.
\eeq
Note that this expression is well defined since the \op \ $\dd$ has no
normalizable \zm s in the right handed spinor subspace.
However in what follows it is convenient that we use the alternative
form
\beq
\psi^{(1)}_L = \dd \frac{1}{D^2} M \frac{1}{\nn} M^+ \psi_{0L}.
\eeq
Due to the antiselfduality property of the anti-\inst \ \f \ we have
implicitly made use of the following identity for the \dop \ in the
anti-\inst \ external \f :
\beq
\dd\,^2(1-\gamma_5) = D^2(1-\gamma_5).
\eeq
It is easy to verify that at each step of the iterative procedure
the
\igty \ \cn \ is trivially satisfied.
The solution is found as an expansion in powers of the \mm \ $M$.
In a closed form it is given by
\beq
\psi_L =\psi_{0L} + \dd\,\frac{1}{D^2-M(1/\nn)M^+\dd}\,M\frac{1}{\nn}\,
M^+ \psi_{0L}, \;\; \eta_R=-\frac{1}{\nn}\,M^+\psi_L.
\eeq
By a
substitution of the expressions (2.21) into (2.16) one can check directly
its validity. Moreover we may note that the operator
$D^2-M(1/\nn)M^+\dd$ has no normalizable zero
modes, at least for values of $M$ not too big,
in order that the expression (2.21) be well
defined.
Indeed the scalar \op \ $-D^2$ is positively definite in the
appropriate space of functions and has no normalizable \zm s.
On the other hand the correction $M (1/\nn) M^+ \dd$
is small at least for small values of the Yukawa couplings and hence of
the matrix $M$.
We therefore may assume that the expression
(2.21) has no singularities at least
for small values of the Yukawa couplings.
Nevertheless we have no arguments against the existence of singularities
for large ones.
Such singularities would probably imply that discrete \zm s appear or
disappear at large Yukawa couplings while the number of normalizable
\zm s is a topological invariant only under 'small' deformations of the
external \f s, and hence of the \mm \ $M$.

Let us now demonstrate that the wave function (2.21) is \nl . For that it
is sufficient to check that expression (2.21) decreases rapidly enough
at large distances $x\to \infty$.
This we can more conveniently do in the regular gauge for the \inst \
configuration.
At infinity we have that
\beq
D_\mu \to U \partial_\mu U^{-1},
M \to U M_0,
\eeq
where $U$ is an element of the $SU(2)$ group that corresponds to the
\inst \ configuration and $M_0$ is a constant matrix.
By a direct substitution of expressions (2.22)
back into eq.(2.21)
we easily find that
\beq
\psi_L \propto U\frac{\partial^2}{\partial^2 - M_0^2} U^{-1} \psi_{0L}.
\eeq
It becomes obvious from the above that
$\psi_{L}$ behaves better at infinity than $\psi_{0L}$
itself
a \nl \ wave function.
Indeed we trivially deduce from eq.(2.23) that
\beq
\psi_{L}\leq U\frac{1}{x^2 M_0^2} U^{-1} \psi_{0L}.
\eeq
Hence $\psi_{L}$ is \nl . Our general arguments are complimentary to
the explicit
construction of the massive fermion \zm \ wave function in
ref.\cite{rub,anselm}.
There for the special case of equal fermionic masses $(h_u=h_d)$
the wave function $(\psi_L,\eta_R)$ in the anti-\inst \ \f \ is found to
be exponentially suppressed at $x\to \infty$ like
\beq
\psi_{L}, \eta_R \propto \frac{e^{-mx}}{x^{3/2}}
\eeq
whereas for the massless \zm \ wave function $\psi_{0L}\propto (x^2 +
\rho^2)^{-3/2}$.
We must emphasize here that our proof of the existence of a \nl \ \zm \
relies heavily on the hypothesis of the behaviour (2.22) of the \mm \
at infinity.
As a consequence the number of \nl \ \zm s is a topological invariant
under smooth deformations of the \mm \ $M$ that preserve
\cn \ (2.22).
Our assumption of (anti)self-duality of the gauge \f \
was necessary for the demonstration
of absence of singularities in expression (2.21).
We could have formally considered cases where the \mm \ does not obey
\cn \ (2.22).
In this case the standard index theorem does
not
hold as we deal with a non-compact $D=4$ manifold. In fact
the external \f s do not rapidly decrease at infinity.
As it has been shown the index theorem for massless fermions should in
this case be generalized in such a way so as to include "phase shifts"
\cite{wipf}.
We believe that such a generalization carries through for the case of
massive
fermions as well. The phase shifts in such a case will receive
contributions both from gauge fields and the \mm .

\section{Massive Fermion in a Sphaleron Field}
\setcounter{equation}{0}

It is well known that the fermionic Hamiltonian in a sphaleron field
possesses a normalizable zero mode \cite{boguta}.
In the standard electro-weak model
where the fermions aquire their masses via the Higgs mechanism
this was also shown to be true for the special case
of degenerate in mass fermion doublets \cite{ringwald}.
In particular by the use of the variational ansatz
the author demonstrated the existence of a normalizable
solution of the zero-energy Dirac equation
in the sphaleron field.
In this section we show the presence of such a massive fermion \zm \ for
the general non-degenerate case.
More specifically we obtain a closed form expression which relates the
massive with the massless \zm .

At first let us consider the case of massless
fermions in an external \sp \ \f .
The $D=3$ \dop \ $\sigma_i D_i$ is antihermitean and purely imaginary
since
\beq
i\sigma_2 \epsilon (\sigma D)^*
i\sigma_2 \epsilon =
- \sigma D,
\eeq
where $\epsilon = i\tau_2$ acts on the $SU(2)$ indices
and $i\sigma_2$ acts on the spinor ones.
Consequently all eigenvalues $i\lambda$ are purely imaginary and all the
wave functions can be chosen real
\beq
i\sigma_2 \epsilon \psi^*_{\lambda} = \psi_{\lambda}.
\eeq
This is in contrast for example to the case of Witten's \an \
where an appropriate anti-Hermitean \dop \ is purely real
\cite{witten}.
There as a consequence its non-zero levels are paired
$(i\lambda,-i\lambda)$
and cross zero only in pairs.
In turn this implies that the number of \zm s of the $D=5$ \dop \
is invariant under smooth deformations of the external \f \ in
agreement with the Atiyah-Singer index theorem
modulo 2 \cite{atiyah}.
This is due to the non-triviality of the fourth
homotopy group $\pi^4(SU(2)) = Z_2$. The situation
is similar for a different reason in the case of the
sphaleron. As this configuration is contractible ($\pi^2(SU(2)) = 0$)
one should not expect the number of \zm s of the
$D=3$ \dop \ to be a topological invariant. Alas this is not the whole
story. The sphaleron configuration (see below) is Parity odd.
The same is true for the Dirac operator in the presence of the sphaleron
gauge field. As a consequence there exists a pairing up of the nonzero
eigenvalues of the Dirac operator $(i\lambda, -i\lambda)$. The number
of normalizable zero modes modulo 2 of the $D=3$ Dirac operator in the
presence of a P-odd external gauge field is a topological invariant
under smooth P-odd deformations of the gauge field. This property also
generalizes for the case of nonvanishing Yukawa interactions as we will
demonstrate shortly.

The \sp \ \con \ in the temporal gauge $A^a_0 = 0$ reads as
\beq
A^a_i = \frac{1}{g} \; \frac{f(\xi)}{x^2} \epsilon_{iaj} x_j,\; \;
\varphi = \frac {v}{\sqrt{2}}
\; \frac{h(\xi)}{\sqrt{x^2}} i(x_i \tau_i)
\left( \begin{array}{c} 0\\ 1
\end{array}  \right),
\eeq
where $\r$ is the Higgs doublet,
$v$ is its vacuum expectation value, $a$ is an $SU(2)$ index,
$g$ is a gauge coupling
constant and $\xi = gv\mid \! x\! \mid$.
The dimensionless functions $f(\xi)$ and $h(\xi)$
have the following asymptotic behaviour:
\beq
f(\xi) = \alpha \xi^2 \;\;\; for \;\;\; \xi \to 0,
\eeq
$$f(\xi) = 1 - \gamma exp(-\xi /2) \;\;\;  for \;\;\; \xi \to \infty ,$$
$$h(\xi) = \beta \xi \;\;\;  for \;\;\; \xi \to 0,$$
$$h(\xi) = 1 - \frac{\delta}{\xi} exp(-\sqrt{2\lambda /g^2 \xi})
\;\;\;  for \;\;\; \xi \to \infty.$$
Here $\lambda$ is the Higgs self coupling constant,
and $\alpha ,\beta ,\gamma$
and $\delta$
are fixed numbers.

The $D=3$ \dop \ in this external
\f \ has a normalizable \zm \ \cite{km}
\beq
(\sigma D) \psi_0 = 0,
\eeq
where
\beq
\psi_0^{\alpha i} = \epsilon_{\alpha i} u(x^2).
\eeq
Here $\alpha$ is a spinor index with $i$ being an $SU(2)$ doublet index.
The function
$u$ has the following behaviour
\beq
u(x) \propto exp(-\alpha x^2) \;\;\; for \;\;\; \xi \to o,
\eeq
$$ u(x^2) \propto 1/x^2 \;\;\; for \;\;\; \xi \to \infty .$$
Let us now consider the spectrum of $D=3$ fermions in the \sp \ \f \
in the presence of Yukawa couplings.
The zero energy equations have the same form as in
the case of the fermionic zero mode in the weak
instanton configuration
\beq
\sigma_i D_i \psi  = \;-M\eta, \qquad  \sigma_i\partial_i
\eta = \; -M^+\psi .
\eeq
We use the notations of sect.2 with the appropriate
dimensional reduction.
Here $\psi$
is a two component Weyl spinor.
By eliminating $\eta_R$ in eq.(3.8) we get
\beq
(\sigma D) \psi = M (\sigma \partial)^{-1} M^+ \psi.
\eeq
Here it should be noted that as $M$ is completely general the \op \
$(\sigma D) - M (\sigma \partial)^{-1} M^+$
does not satisfy relation (3.1) and therefore is not purely imaginary.
This is true only in the "symmetric" case where $h_u = h_d$.
There all eigenfunctions can be chosen to be real just as in eq.(3.2).
In this limit the approximate solution of eq.(3.8) is possible as it was
actually done in ref.\cite{ringwald}.
In our approach the Yukawa couplings are taken to be arbitrary.
We may note that the operator $\sigma D-M(\sigma \partial)^{-1} M^+$,as
well as the massless Dirac operator, is P-odd. Hence the number of its
zero modes is an invariant modulo 2 under smooth deformations of the
gauge and Higgs fields which preserve their P-odd character. Since the
massless Dirac operator has exactly one normalizable zero mode , this is
also the case for at least small values of the Yukawa couplings. Below
we give an explicit construction of a fermionic zero mode for
nonvanishing Yukawa interactions.
We now try to solve eq.(3.9)
iteratively starting with $\psi_0$ as our zero level approximation.
The solution $\psi$ is in this sense expanded in powers of the \mm \
$M$.
At the first step of the iteration procedure one gets
\beq
(\sigma D) \psi^{(1)} = M (\sigma \partial)^{-1} M^+ \psi_0.
\eeq
Let us firstly check the \igty \ of the above equation.
By this we mean that
\beq
a_0 = \int d^3 x  \psi^+_0 M (\sigma \partial)^{-1} M^+ \psi_0 = 0.
\eeq
It is easy to see that this is true.
Indeed we note that the \sp \ \con \ (3.3) is parity
$(P)$ odd, i.e. under the transformation $x_i \to -x_i$.
The integrand therefore in eq.(3.11) is $P$-odd
and hence the integral over all directions of $x_i$ vanishes.
Actually one should be careful here since the integral (3.11)
is formally logarithmically divergent.
In order to make our claims more rigorous we introduce a dimensional
regularization.
We do our computation in $D = 3 - \epsilon$ dimensions and then take the
limit $\epsilon \to 0$.
Our final expressions in eq.(3.11)
are $\epsilon$ independent in this limit.

We must emphasize at this point that
we can write down the intergrability \cn \ for the exact solution to
eq.(3.9).
By noting that the left hand side of eq.(3.9) is orthogonal to $\psi_0$
we get
\beq
\int d^3 x  \psi^+_0 M (\sigma \partial)^{-1} M^+ \psi = 0.
\eeq
It is worthy to emphasize at this point that
there is no problem of \igty \ in the case
of the $D=4$ \dop \ \cite{anselm} in the weak \inst \ \con .
There \igty \ is automatic as a consequence of the anti-selfduality
of the anti-\inst \ gauge
\f .
We find the exact solution to eq.(3.9) to be given by
\beq
\psi = \psi_0 + (\sigma D) \frac{1}{((\sigma D)^2 + \alpha P -
M (\sigma \partial)^{-1} M^+ (\sigma D))} \\
M \frac{1}{(\sigma \partial)} M^+ \psi_0.
\eeq
$P$ is the projector onto the \zm \ subspace and $\alpha$ is a
regularizing parameter which is introduced here to make the expression
well defined.
One can straightforwardly check that our solution is independent
of $\alpha$ and obeys eq.(3.9).
To this end
we act on the wave
function given by (3.13) by the \op \ $(\sigma D) - M(1/(\sigma\partial)) M^+$.
We find an expression that is proportional to the following one
\beq
P \frac{1}{(\sigma D)^2 + \alpha P -
M (\sigma \partial)^{-1} M^+ \sigma D}
M \frac{1}{(\sigma\partial)} M^+
\psi_0 =0,
\eeq
where $P = \psi_0 \psi^+_0$.
In order to compute the above expression it is convenient that we
explicitly split the associated functional space into
two subspaces $F = F_{\bot} \oplus F_0$,
where $F_0$ is the one dimesional one which is generated by $\psi_0$
while $F_{\bot}$ is orthogonal to $\psi_0$.
The non-local
\op \ in eq.(3.14) can be represented as a block matrix.
More graphically
\beq
((\sigma D)^2 + \alpha P -
M (\sigma \partial)^{-1} M^+ (\sigma D))^{-1} =
\left( \begin{array}{cc} A & 0 \\ B & C \end{array} \right),
\eeq
where the \op \ $A=((\sigma D)^2 -
(1-P)M (\sigma \partial)^{-1} M^+ (\sigma D))^{-1}$ maps $F_{\bot}$
into  $F_{\bot}$,
$B=(1/\alpha)
P M (\sigma \partial)^{-1} M^+ (\sigma D) A$
maps $F_{\bot}$
into $F_0$ and finally $C=(1/\alpha)P$ maps $F_0$
onto  $F_0$.
In such a formulation of the problem we have
\beq
P = \left( \begin{array}{cc} 0 & 0 \\ 0 & 1 \end{array} \right)
\eeq
while $M (\sigma \partial)^{-1} M^+ \psi_0$
is now identified by the vector
\beq
((1-P) M (\sigma\partial )^{-1} M^+ \psi_0,
P M (\sigma\partial )^{-1} M^+ \psi_0 ).
\eeq
By a direct substitution of eqs.(3.15) and (3.16) into eq.(3.14) we
easily get
\begin{eqnarray}
PM\frac{1}{(\sigma\partial)}M^+ \psi_0 +
PM\frac{1}{(\sigma\partial)}
M^+ \sigma D \frac{1}{(\sigma D)^2 -
M (\sigma \partial)^{-1} M^+ (\sigma D)} \times \nonumber \\
\times (1-P)M\frac{1}{(\sigma\partial)} M^+ \psi_0 =0 \;\;\;
\end{eqnarray}
Equivalently
\begin{eqnarray}
\int d^3 x
\psi_0^+ \left( M \frac{1}{(\sigma\partial)} M^+ +
M \frac{1}{(\sigma \partial)} M^+
\frac{1}{(\sigma D) -
(1-P) M (\sigma \partial)^{-1} M^+ (1-P)}\times \right. \nonumber \\
\left. \times M \frac{1}{(\sigma\partial)} M^+\right)
\psi_0 =0. \;\;\;\;\;
\end{eqnarray}
Note that this expression is well defined since
$M (\sigma\partial)^{-1} M^+
\psi_0$
is orhogonal to $\psi_0$ due to eq.(3.11).
This integral is a modification of the identity (3.11).
It vanishes as a result of the $P$-oddness property of the external \f
s.
However we must show that this solution obeys the \igty \ \cn \ (3.12).
By using the block matrix representation which is given above,
it is straightforward to show that the \cn \ (3.12) is in fact
equivalent to the identity (3.18).
Finally we like to check that the wave function (3.13) is \nl .
The argument here is similar to the \inst \ one ($D=4$).
At spatial infinity the gauge \f \ is a pure gauge and along
with
the \mm \ $M$ are given by
\beq
A_i = U \partial_i U^+, \; M = M_0 U.
\eeq
Here $i = 1,2,3$ and $M_0$ is a constant,
whereas $U$ is an element of $SU(2)$.
For large distances the normalization integral is given by
\beq
\int d^3 x \psi^+ \psi \propto
\int d^3 x \xi_0^+ \partial^4 /(\partial^2 - M_0^2)^2 \xi_0,
\eeq
where $\xi_0 = U M_0$.
As this expression is manifestly finite the massive \zm \ $\psi$
is \nl \ if we take into account eq.(3.6) for the massless \zm .
At this point we believe that
our arguments with regard to the existence of a
massive fermion \zm \ in the background of a  \sp \
\f \ become complete.
Level crossing is hence automatically obtained for the general case of
fermions getting their masses from Yukawa interactions.

\section{Global Anomaly for Massive Fermions}
\setcounter{equation}{0}

It is well known that $SP(n)$ gauge theories have no
local anomalies. Moreover when coupled with an odd
number of Weyl fermions they possess
a global anomaly and become self inconsistent.
The simplest such example is an $SU(2)$ gauge
theory with one fermionic Weyl doublet.
This model was previously considered
in some detail for the case of massless fermions \cite{witten}.
In this section we extent Witten's arguments in order to take into
account the presence of Yukawa interactions.
We will show that the Atiyah-Singer index theorem
mod 2 is sufficient for the presence of the \an \ as a consequence
of the existence of a massive normalizable \zm \ of the $D=5$ \dop .

Let us first sketch Witten's arguments \cite{witten}.
The partition function of
the euclidean version of the model of a doublet of
massless fermions coupled to an
$SU(2)$ gauge field reads as follows
\beq
Z = \int D\psi_L D\psi^+_L \int DA_{\mu} exp(-\int d^4 x
[(1/2g^2)Tr F_{\mu\nu}^2 + \psi^+_L i\dd \psi_L]).
\eeq
There  $A_{\mu}$ is an $SU(2)$
gauge field, $\psi_L$ is a left-handed Weyl fermion
doublet,
$g$ is the gauge coupling constant, $\dd\ = D_{\mu}\gamma_{\mu}$
is a \dop \ restricted to act on a Weyl doublet.
The fermionic
part of the integral eq.(4.1) is ill defined.
However it can be formally integrated as the square root of
a functional integral over Dirac fermions.
As such it implies the doubling of the
fermionic degrees of freedom
from one to two Weyl left-handed doublets.
Because the $1/2$ representation of $SU(2)$ is pseudoreal a left-handed
doublet can be mapped to the right-handed one.
A theory with two left-handed Weyl doublets
is thus equivalent to a vector-like one with a single Dirac doublet.
The fermionic functional integral is given by $det(i\dd)$
and it is well defined.
Then we formally have that
\beq
\int D\psi_L D\psi^+_L exp \int \psi^+_L i\dd\psi_L =
(det i\dd)^{1/2}.
\eeq
The sign of the
square root is ill defined. As a way out Witten defines
the root in eq.(4.2) as the product of all positive eigenvalues of
a \dop .

If for a given $A_{\mu}$ the sign in eq.(4.2)
is arbitrarily fixed as Witten showed
there always exists
a configuration $A_{\mu}^U$ that
can be reached continuously
from $A_{\mu}$ for which the fermionic determinant has an opposite
sign, i.e.
\beq
(det i\dd (A_{\mu}))^{1/2} = - (det i\dd (A_{\mu}^U))^{1/2}.
\eeq
Here $A_{\mu}$ is taken to be the gauge transformed
\con \ of $A_{\mu}$.
This means that the partition function
\beq
Z = \int DA_{\mu} (det i\dd )^{1/2} exp (-(1/2g^2)\int d^4 x Tr F^2_{\mu\nu})
\eeq
vanishes due to the contribution with an opposite sign of $A_{\mu}$ and
$A_{\mu}^U$.
This occurs in all topologically distinct sectors of
the field \con s $A_{\mu}$ independently.

As we continuously vary
the external field value from $A_{\mu}$ to $A_{\mu}^U$ an odd number of these
eigenvalues flow through zero switching their sign.
Such a \lc \ effect is a reflection of the existence of an odd number of
\nl \ \zm s for a properly defined $D=5$ \dop \ in a topologically
nontrivial gauge \f .
It is a result of the nontrivial fourth homotopy group of $SU(2)$
\beq
\pi^4(SU(2)) =Z_2.
\eeq
The D=5 topologically non-trivial gauge field must accordingly
belong to the
nontrivial homotopy class in $Z_2$
and it is the interpolating \con \ between $A_{\mu}$ and $A_{\mu}^U$.
An appropriate $D=5$ \dop \ is actually a generalization of the $D=4$
one that includes evolution in the fifth coordinate $\; t$.
In what follows we generalize Witten's argument for the case of Yukawa
interactions and the presence of fermion mass through the Higgs
mechanism.
In this context we will prove Witten's conjecture of the validity
of his arguments for this case too.
It is worthwhile to also stress that the $SU(2)$ \gl \ \an \
can also be understood as a manifestation of the existence of a local
\an \ for an $SU(3)$ gauge theory \cite{klinkhamer}.
{}From this point of view the generalization to the case of massive
fermions is of course straightforward.
This is due to the fact that the local \an \ is independent of Yukawa
couplings which are $SU(3)$ invariant \cite{fuj}.
It seems
nevertheless interesting to try to understand the issue in terms of
a level crossing phenomenon
for the $D=4$ Dirac \op .
This is the aim of what is to follow.
We do it by proving a generalization of the index theorem
mod 2 for the massive $D=5$ \dop .

We first consider the index theorem
mod 2 for a massless $D=5$ Dirac \op .
Instead of making
a unitary transformation to a real representation
for the $D=5$ fermions which transform under the
$O(4) \times SU(2)$  group \cite{witten} we introduce
\beq
D\hspace{-.65em}/ = (\gamma_5 \nabla_t + \nabla \hspace{-.65em}/ ).
\eeq
Here $\nabla_{\mu}$ is the usual covariant derivative,
$\nabla \hspace{-.65em}/ = \nabla_{\mu} \gamma_{\mu}$,
whereas $\nabla_t$
is the covariant derivative for the
fifth coordinate.
Eigenvalues and eigenfunctions are defined by the following equation
\beq
D\hspace{-.65em}/  \psi = i \lambda \psi ,
\eeq
where $\lambda$'s are real since $\dd$ is antihermitean.
The \op \
$D\hspace{-.65em}/ $
is real in the following sense
\beq
C\epsilon \gamma_5 \dd^* C\epsilon \gamma_5 = \dd  .
\eeq
Here $C=i\gamma_2\gamma_0$ and $C\gamma_5$
are the usual $D=4$ and $D=5$ charge conjugation matrices,
$\epsilon$ is a $2\times 2$ antisymmetric matrix that acts on $SU(2)$
indices.
The reality \cn \ implies the pairing up of all the non-zero
eigenvalues.
More precisely
if $\dd \psi = i \lambda \psi$ then
\beq
\dd (C\epsilon \gamma_5 \psi^*) = -i \lambda C\epsilon \gamma_5 \psi^.
\eeq
The zero mode wave
functions $\psi_0$ (if they exist) can be chosen real
satisfying a Majorana-like \cn :
\beq
C\epsilon \gamma_5 \psi_0^* = \psi_0.
\eeq
This reality condition means that the number of zero modes
is a topological invariant modulo 2 since non-zero modes can
cross the zero level only in pairs when external fields vary smoothly.
The crucial point is that such a zero mode does exist in the external
field due to the index theorem modulo 2 \cite{atiyah} if
the external \f \ is topologically nontrivial.
The reality of the Dirac operator
$\dd$ guarantees
a pairing of its eigenvalues.
They correspond to the existence of the nontrivial homotopy class that
arises in the nontrivial topology $\pi^4 (SU(2)) = Z_2$.

We now proceed to examine massive fermions in such an external field.
In this model we have a left handed chiral fermion
$SU(2)$ doublet $q_L$ and a pair of right handed singlet fermions
which can be combined into a doublet $q_R$.
This is necessary for the introduction of fermionic masses.
The problem at hand now is how to generalize the definition of the
chiral fermionic determinant for the massive case.
We find it convenient and natural to define it as a square root of the
fermionic functional integral for two fermionic $(q_L^i,q_R^i)$,
$i=1,2$, multiplets with the same \mm \ $M$.
We follow, to this end, the recipe of sect.2.
We combine them into the Dirac spinors $\psi$ and $\eta$
whereas $\psi$ is an $SU(2)$ doublet and $\eta$ is a couple of $SU(2)$
singlet spinors.
As a result we get the \op \ in the fermionic kinetic term
as before for the particular case where $M=M'$.
This \op \ acts in the space of pairs $(\psi ,\eta)$ and reads as
\beq
T(M) = \left( \begin{array}{cc}  \nabla\hspace{-.75em}/ &
MR - \epsilon M^* \epsilon L\\
M^+ L - \epsilon M^T \epsilon R & \partial\hspace{-.55em}/
\end{array} \right),
\eeq
where $L(R) = (1+(-)\gamma_5 )/2$.
We restrict ourselves to the case of equal fermion masses
$(h_u=h_d)$.
In this case $M=-\epsilon M^* \epsilon$ with $M$ being
proportional to an element of $SU(2)$ group.
This assumption will appear necessary with regard to a definition of an
appropriate chiral fermionic determinant and a $D=5$ antihermitian
euclidean \dop .
Indeed we note that $PT(M=-\epsilon M^*\epsilon)$ is
antihermitian where $P$ is given by
$$P =
\left( \begin{array}{cc} 1 & 0\\ 0 & 1
\end{array} \right)$$.
We thus reach to the following definition for the chiral fermionic
determinant for the massive case
\beq
\Delta_{ch} = (det \; T(M))^{1/2}.
\eeq
In order to generalize our discussion of \lc \ we now define an
appropriate $D=5$ \op \ (for $M=-\epsilon M^*\epsilon$) as
\beq
\hat{D} =
\gamma_5 \left( \begin{array}{cc}  \nabla_t  &  0\\ 0 &
-\partial_t \end{array} \right) +P\hat{T} =
\left( \begin{array}{cc} \gamma_5 \nabla_t + \nabla\hspace{-.75em}/ &
M\\
-M^+ &
-\gamma_5 \partial_t - \partial\hspace{-.55em}/ \end{array}
\right).
\eeq
We observe that $\dd$ satisfies the following reality condition
\beq
C\epsilon \gamma_5 \hat{D}^* C \epsilon \gamma_5 = \hat{D}.
\eeq
This is an important property that our \dop \ ($D=5$) must satisfy in
complete analogy with the massless fermion case.
In effect there is a pairing up of all of the non-zero eigenvalues
in its spectrum.
This furthermore implies that the number of its zero modes, if they exist,
is a topological invariant mod 2.

The solution $\Psi_0$ to the \zm \ equation
\beq
\hat{D}\Psi_0 = 0
\eeq
can be chosen real such that
\beq
C\epsilon \gamma_5 \Psi_0^* = \Psi_0.
\eeq
Notice that such a solution is not of any
definite chirality.
We may now represent $\Psi_0$ as a superposition of
an $SU(2)$ doublet
$\psi$ and two $SU(2)$ singlets $\chi_1$ and $\chi_2$, i.e.
$\Psi_0 = (\psi, \chi)$ with $\chi = (\chi_1 ,\chi_2)$.
The zero mode equations then read as
\beq
\dd \psi + M\chi = 0, \qquad
M^+ \psi + \dd_0 \chi = 0,
\eeq
where
\beq
(\gamma_5 \nabla_t + \nabla\hspace{-.65em}/ ) = \dd, \qquad
(\gamma_5 \partial_t + \nn ) = \dd_0.
\eeq
By eliminating $\chi$ from eqs.(4.17)
we get
\beq
\dd \psi = M (1/ \dd_0) M^+ \psi.
\eeq
Let us assume that there is exactly one \zm \ of the \op \ $\dd$
such as in the case of eq.(4.15).
We want to find the massive \zm \ wave function,
and moreover express the solution of the above equation
in terms of the \zm \ wave function $\psi_0$
for the massless case. This state of affairs is similar to that of the
$D=3$ \sp \ example.

We search for a solution iteratively by starting with $\psi_0$
as a zero
level approximation.
The first level correction satisfies the following equation
\beq
\dd \psi^{(1)} = M \frac{1}{\dd_0} M^+ \psi_0 .
\eeq
The integrability \cn \ is deduced from the orthogonality
of the left hand side of eq.(4.20) to $\psi_0$
\beq
a_0 = \int d^5 x  \psi^+_0 M \frac{1}{\dd_0} M^+ \psi_0 = 0.
\eeq
Indeed if we make a transposition in eq.(4.21) and susequently take into
account the reality \cn s (4.14,4.16) for the anti-hermitean \op \
$\dd_0$ we get that $a_0 = -a_0 =0$.

It is amusing to note
that we did not use so far any properties of the external \f s in
contrast to the \sp \ case.
This is due to the reality property
of the $D=5$ \op \ $\dd$ whereas the $D=3$
\dop \ is purely imaginary.
We may now solve eq.(4.20)
\beq
\psi^{(1)} = \frac{1}{\dd} M \frac{1}{\dd_0} M^+ \psi_0.
\eeq
This is a well defined expression on the basis of eq.(4.21).
The exact solution to eq.(4.19) is given by
\beq
\psi = \psi_0 + \dd \frac{1}{\dd^2 + \alpha P -
M \dd_0^{-1} M^+ \dd } \\
M \frac{1}{\dd_0} M^+ \psi_0,
\eeq
where $P$ is the \zm \ subspace projector.
This expression is well defined and does not depend on
a regularizing parameter $\alpha$.
By substitution of (4.23) for $\psi$
into eq.(4.19)
we get the following equation
\beq
P \frac{1}{\dd^2 + \alpha P - M \dd_0 M^+ \dd} \; M\frac{1}{\dd_0}
M^+ \psi_0 =0.
\eeq
By using the block
matrix representation similar to the case of the \sp \
we find an eqivalent form of eq.(4.24)
\begin{eqnarray}
\int d^5 x \psi_0^+ \left( M \frac{1}{\dd_0} M^+ +
M \frac{1}{\dd_0} M^+ (1-P)
\frac{1}{\dd  -
(1-P) M \dd_0^{-1} M^+ (1-P)}\times \right. \nonumber \\
\left. \times (1-P) M \frac{1}{\dd_0} M^+\right)
\psi_0 =0.
\end{eqnarray}
The above expression
is well defined.
It is certainly satisfied as a consequence of the reality \cn s
(4.14,4.16) and the mass degeneracy \cn \ $M = -\epsilon M^* \epsilon$.
This can be checked by a transposition similar to the case of eq.(4.21).

Let us now turn to the issue of \nty \ of $\psi$ in eq.(4.23).
It is easy to see that this wave function decreases more rapidly at
infinity provided that the \mm \ $M$ is an asymptotically covariant
constant similar to the case of the $D=4$
\inst \ $+$ Higgs \con s.
This necessitates that the gauge \f \ is asymptotically a pure gauge
$A_{\mu} \propto U
\partial_{\mu} U^{-1}$ with the \mm \ being $M \propto UM_0$
at infinity where $U$ is an element of the $SU(2)$ group and $M_0$ is a
constant.
In this case the wave function (4.23) is normalizable.
It is to be emphasized at this point that by changing $M$ from zero to
a nonzero value we can obtain additional zero modes but always in pairs.
It is of some significance to note that this \cn \ implies that
in the presence of Yukawa interactions the interpolation between
two $D=4$ \con s of gauge fields $A_{\mu}$ and $A^U_{\mu}$
which is given by a $D=5$ gauge \f \ \con \ should also include
Higgs \f s.

We now turn to the case where the \op \ $\dd$ has a number of
normalizable zero modes $\psi_{0i}$, $i=1,...,N$.
These zero modes can be chosen to obey the reality condition
\beq
C \gamma_5 \epsilon \psi_{0i}^* = \psi_{0i}.
\eeq
In order to solve eq.(4.19) we start the
iteration procedure with a linear combination
\beq
\psi_0 = c_i \psi_{0i}
\eeq
as a zero level approximation.
At the first step we get eq.(4.20).
The integrability condition reads now as
follows
for all $i$
\beq
\int d^5 x \psi^+_{0i} M \dd_0^{-1} M^+ \psi_0 = 0.
\eeq
This means that the vector $c_i$ should be annihilated by the matrix
\beq
a^{(0)}_{ij} = \int d^5 x \psi^+_{0i} M\dd_0^{-1} M^+ \psi_{0j}.
\eeq
A solution of eq.(4.20) and hence of eq.(4.19)
exists only if the matrix $a^{(0)_{ij}}$
has zero eigenvalues.
By considering the Hermitian conjugate and the complex one of $a_{ij}$
we can easily check
that this matrix is anti-hermitian and purely real
\beq
a^{(0)+}_{ij} = -a^{(0)}_{ji},\;\;\;
a^{(0)*}_{ij} = a^{(0)}_{ij}.
\eeq
In general this matrix is non-zero and belongs to the $O(N)$ algebra.
Let us  first consider the case of an odd number of $\psi_0$'s, i.e.
$N=2n+1$.
It is easy to see that the matrix $a^{(0)}_{ij}$ has at least one zero
eigenvalue. Indeed it is clear that $O(N)$
transformations preserve the reality
conditions (4.26) and (4.16).
By making successive $O(N)$ rotations we can bring the matrix
$a^{(0)}_{ij}$
in the form for which all non-vanishing
elements are organized into cells proportional to
\beq
\left( \begin{array}{cc} 0 & 1\\ -1 & 0 \end{array} \right)
\eeq
which are located along the diagonal.
These cells act onto $n$ orthogonal subspaces of the space of $O(2n+1)$
vectors.
Since $N$ is odd at least one of eigenvalues of the matrix
$a^{(0)}_{ij}$ is zero. The other eigenvalues are combined into pairs
$(i\lambda,-i\lambda)$ corresponding to cells given above.
In general they do not vanish. However when the mass matrix
$M$ varies
the zero eigenvalues of the matrix
$a^{(0)}_{ij}$ can appear or disappear
in pairs.

In the case of even $N$, i.e. $N=2n$,
all eigenvalues of $a^{(0)}_{ij}$ are generically non-zero and
are arranged in
pairs $(i\lambda, -i\lambda)$ as above.

We get the zero level approximate
solution to eq.(4.19) by picking the eigenvector of $a^{(0)}_{ij}$ that
corresponds to the zero eigenvalue for the vector $c_i$ in eq.(4.27).
It is obvious from the above construction that the
'oddness' of the number of the fermionic
\zm s does not change.
In particular there is at least one normalizable zero mode
of massive fermion in the $D=5$ topologically nontrivial
configuration since
it has exactly one zero mode for massless ones.

The exact solution to eq.(4.19) is given by (4.23) where $P$ is the
projector onto the total \zm \ subspace.
We must be more careful here than with the case of the one massless \zm \
$(N=1)$ since in general the function $M \dd_0^{-1} M^+ \psi_0$ is not
orthogonal to the \zm \ subspace.
By the use of the block matrix representation it is easy to see that the
expression (4.23) obeys eq.(4.19) provided that the following holds true
for all $i$
\begin{eqnarray}
\int d^5 x \psi_{0i}^+ \left( M \frac{1}{\dd_0} M^+ +
M \frac{1}{\dd_0} M^+ (1-P)\times \;\;\; \right. \\
\left. \times \frac{1}{\dd  -
(1-P) M \dd_0^{-1} M^+ (1-P)}
(1-P) M \frac{1}{\dd_0} M^+\right)
\psi_0 =0. \;\; \; \nonumber
\end{eqnarray}
The above expression is well defined and implies that the vector
$c_j$ should be the eigenvector that corresponds to the zero
eigenvalue of the following matrix
\begin{eqnarray}
a_{ij} = \int d^5 x \psi_{0i}^+ \left( M \frac{1}{\dd_0} M^+ +
M \frac{1}{\dd_0} M^+ (1-P)\times \;\;\; \right. \\
\left. \times \frac{1}{\dd  -
(1-P) M \dd_0^{-1} M^+ (1-P)}
(1-P) M \frac{1}{\dd_0} M^+\right)
\psi_{0j}, \nonumber \;\;\;
\end{eqnarray}
i.e.
\beq
\sum_j \; a_{ij}\; c_j = 0.
\eeq
In the leading order in the Yukawa
couplings this \cn \ reproduces eq. (4.28).
It can be easily seen that it is equivalent to the \igty \
\cn \ analogous to eq.(4.21).
One may observe that the matrix $a_{ij}$ is purely real and
antisymmetric by using the reality \cn \ (4.26).
By repeating here the procedure we followed for the leading
approximation we find that the matrix $a_{ij}$
has exactly an odd number of zero eigenvalues for odd $N=2n+1$
and correspondingly an even number of zero eigenvalues for even
$N=2n$.
In the latter case the zero eigenvalues can be absent.
Hence there is an invariance for the odd/even number of fermionic
zero modes under smooth deformations of the \mm \ $M$ (see eq.(4.19)).

At this point our arguments that generalize Witten's observation of
an $SU(2)$ \an \ for odd number of massless fermion doublets in the
presence of Yukawa
interactions are complete.
The existence of the \zm \ for massive $D=5$ \dop \ was demonstrated
for the case of degenerate \mm .
The treatment of the general non-symmetric $(h_u \neq h_d)$ case
will be present elsewhere.
\subsection*{Conclusions}
In the present work we considered the effect of Yukawa couplings and
fermion masses on the problem of existence of fermionic zero modes in
the presence of topologically nontrivial fields for an $ SU(2)$ gauge
theory. We demonstrated the existence of normalizable zero energy
solutions of the Dirac operator in the presence of a sphaleron the
instanton and that of the $SU(2)$ global anomaly. As such we
equivalently confirmed the existence of the fermionic energy level
crossing phenomenon in the presence of time dependent external gauge and
Higgs fields in all three physical situations. While the first two cases
are relevant to the Baryon and Lepton quantum number violations in the
standard electroweak theory the latter is associated with the global
$SU(2)$ anomaly in the presence of an odd number of Weyl doublets.

\subsection*{Acknowledgments}
One of us A.J. acknowledges the high energy group at NBI for its
hospitality. The present work was supported in part by a NATO grant
GRG 930395.


\begin{thebibliography}{11}
\bibitem{sc} S.Coleman. Aspects of Symmetry. Cambridge University Press.
\bibitem{hooft} G.'t Hooft. Phys.Rev.D 14(1976)3432; Phys.Rev.Lett.
37(1976)8; Phys. Rev.D
18(1978)2199.
\bibitem{callan} C.G.Callan,  R.F.Dashen, D.J.Gross.
Phys.Rev.D 17(1978)2717.
\bibitem{kiskis} J.Kiskis. Phys.Rev.D 18(1978)3690.
\bibitem{amb} J.Ambjorn, J.Greensite, C.Petersen. Nucl.Phys.B
221(1983)381. H.B.Nielsen and A.Wirzba.Les Houches,March 24-April 2
1987. Berlin, Springer Proc. in Physics.26(1988); H.B.Nielsen and
M.Ninomiya. Int. J. of Mod. Phys.A, vol 6, No 16(1991) 2913 and
references therein.
\bibitem{christ} N.Christ. Phys.Rev.D 21(1980)1591.
\bibitem{rubakov} V.A.Rubakov. Nucl.Phys.B 256(1985)509.
\bibitem{boguta} J.Boguta, J.Kunz. Phys.Lett.B 154(1985)407.
\bibitem{ringwald} A.Ringwald. Phys.Lett.B 213(1988)61.
\bibitem{witten} E.Witten.Phys.Lett. 117B(1982)324.
\bibitem{rub} N.V.Krasnikov, V.A.Rubakov, V.F.Tokarev. J. of Phys.
A12(1979)L343.
\bibitem{anselm} A.A.Anselm, A.A.Johansen. Preprint
LNPI-1778, March 1992,to appear in Nucl.Phys.B (1993).
\bibitem{km} F.Klinkhamer, N.Manton.Phys. Rev.D 30(1984)2212.
\bibitem{hu} S.T.Hu. Homotopy Theory. Academic Press, New York, 1959,
ch.11.
\bibitem{wipf} P.Forgacs, L.O'Raifeartaigh, A.Wipf.
Nucl.Phys.B 293(1987)559.
\bibitem{atiyah}M.F.Atiyah, I.M.Zinger, Ann. of Math. 93(1971)119.
\bibitem{nair}S.Elitzur, V.P.Nair. Nucl.Phys.B 243(1984)205.
\bibitem{klinkhamer} F.R.Klinkhamer. Phys.Lett.B 256(1991)41.
\bibitem{fuj} K.Fujikawa. Phys.Rev.D 29(1984)285.

\end{thebibliography}
\end{document}